# On the structure of MHD shock waves in a viscous gas

**R. K. Anand and Harish C. Yadav**

**Department of Physics, University of Allahabad, Allahabad-211002, India**

**e-mail: anand.rajkumar@rediffmail.com**

**Abstract** The exact solutions for MHD shock waves in an ideal gas are obtained taking into consideration only the viscosity of the gas. In view of an axial magnetic field, the analytical expressions for the particle velocity, temperature, pressure and change-in-entropy within the shock transition region are obtained. The flow variables are numerical analysed to explore the influence of static magnetic field, shock strength, specific heat ratio, initial pressure, initial density and coefficient of viscosity on the flow variables. The findings confirm that thickness of MHD shock front increases with increase in the viscosity of the gas and the change in thickness is more noticeable for large values of the strength of magnetic field. The results provided a clear picture of whether and how the viscosity of gas and the magnetic field affect the thickness of shock front.

## 1. Introduction

Magnetohydrodynamic (MHD) shock waves are frequently encountered in various astrophysical and geophysical phenomena. A better understanding of the mechanisms governing emergence and evolution of disturbances in a viscous MHD is essential for the development of efficient methods for controlling the turbulent transition in the flow around the hypersonic flying objects (at lower Mach number). Among the industrial applications involving applied external magnetic fields are drag reduction in duct flows, design of efficient coolant blankets in tokomak fusion reactors, control of turbulence of immersed jets in the steel casting process and advanced propulsion and flow control schemes for hypersonic vehicles and missiles.

Studies of the structure of shock waves in an ideal gas have been extensively reported in the literature. Early theoretical investigations were restricted to the perfect gas equation of state and the Navier-Stokes relations with constant coefficients of heat conduction and viscosity. Rankine [1] published his dissertation on the structure of shock waves in 1870 in which he gave an explicit solution for the case of heat conduction only. Hugoniot's work [2] was published in 1889. Their equations for shock jumps in particle velocity, stress, and specific internal energy have become known as the Rankine-Hugoniot conditions. Later, an explicit solution for viscosity alone was given by Taylor [3], who also gave an explicit formula valid for weak shock waves with both viscosity and heat conduction present. The solutions were derived from the conservation laws of mass, momentum and energy under the assumption that the shock is single steady wave of finite thickness. A conclusion was that, for a weak shock wave, the thickness becomes very large, varying as the reciprocal of the shock strength. Hoffmann and Teller [4] extended the Rankine-Hugoniot conditions of classical hydrodynamics to shock waves in an infinitely conducting fluid with superposed magnetic field. The mathematical discontinuity in the physical variables given by the Rankine-Hugoniot conditions at a shock front is, however, not physically possible, and it is well known that considerations of dissipation of energy by viscosity and heat conductivity enable the physical quantities to vary continuously and result in a finite width of the shock front. Applying similar considerations Sen [5] described the structure of a magnetohydrodynamic shock wave in infinitely conducting plasma (a macroscopically neutral, ionized gas). Richtmyer and Von Neumann [6] suggested a method which avoids such difficulties, particularly for numerical calculation. They observed that the addition of a particular viscosity like term into gas dynamics equations could lead to the continuous shock flow in which the finite thickness of discontinuities at the shock wave was removed and replaced by a region in which physical parameters changed rapidly and smoothly. The thickness of a thin transition layer representing the shock across which the gas undergoes transition from the initial to the final state, and this layer is generally known as a shock front. In this layer the density, the pressure and the velocity of fluid change as entropy increases. The increase in entropy indicates that there is a dissipation of mechanical energy and thus an irreversible conversion of mechanical energy into heat energy takes place in the transition layer. The dissipative processes of viscosity (internal friction) and heat conduction are attributable to the molecular structure of a fluid. Such processes create an additional, non-hydrodynamic transfer of momentum and energy, and result in non-adiabatic flow and in the thermodynamically irreversible transformation of mechanical energy into heat. Viscosity and heat conduction appear only when there are large gradients in the flow variables within a shock front. Zel'dovich and Raizer [7] studied the entropy production within a planar viscous shock wave and gave an analytical model for the shock process with effects of viscosity and heat conduction based on Huguenot curves. Landau and Lifshitz [8] investigated the weak viscous shock waves with respect to the small changes in the flow variables. Painter [9] studied a viscous shock wave in an elastic tube. Maslov [10] studied the wave processes in a viscous shock layer and control of fluctuations. The authors [11] investigated the propagation of planar, cylindrically and spherically viscous shock waves in an ideal gas. Recently, Anand [12] formulated the shock jump relations for MHD shock waves in non-ideal gas and discussed the change-in-entropy across the shock front.

The main purpose of this paper is to present an exact solution for one dimensional MHD shock wave in an ideal gas, by assuming that the viscosity is present, and that the heat conductivity is neglected, to find the effect of viscosity on the shock wave profile. For this purpose, a model was developed to provide a simplified, complete treatment of the structure of planar MHD shock waves in ideal gas. The general non-dimensional forms of the analytical expressions for the distribution of particle velocity, temperature, pressure and change-in-entropy within the shock transition region are derived, assuming the medium to be viscid, non-heat conducting, electrically infinitely conducting, initially uniform and at rest. The magnetic field is assumed having only constant axial component which is perpendicular to the shock front. The numerical estimations of flow variables are carried out using MATLAB code. The effects of viscosity are investigated on the MHD shock transition region. This model appropriately makes obvious the effects due to an increase in (i) the propagation distance from the centre of front, (ii) the strength of magnetic field, (iii) the strength of shock wave i.e., Mach number, (iv) coefficient of viscosity and (v) the adiabatic index, on the particle velocity, temperature, pressure and change-in-entropy within the shock transition region.

The rest of the paper is organized as follows. Section 2 describes the general assumptions and notations, the equations of motion and boundary condition. In section 3 the analytical expressions for flow variables are obtained.

A brief discussion with results is presented in section 4. The findings are presented in section 5 with details on which effects were accounted for and which were not.

**2. Basic equations and boundary condition**

The conservation equations governing the flow of a one-dimensional, viscous, ideal gas under an equilibrium condition in presence of a magnetic field can be expressed conveniently in Eularian coordinates as follows:

$$\frac{\partial \rho}{\partial t}+u\frac{\partial \rho}{\partial r}+\rho\frac{\partial u}{\partial r}=0 \ , \tag{1}$$

$$\frac{\partial(\rho u)}{\partial t}+\frac{\partial(\rho u^2+P-q)}{\partial r}+\frac{\mu'}{2}\frac{\partial H^2}{\partial r}=0 \, , \tag{2}$$

$$\frac{\partial[\rho E+\rho u^2/2]}{\partial t}+\frac{\partial[\rho\, u(E+u^2/2)+pu-qu]}{\partial r}=0 \ , \tag{3}$$

$$\frac{\partial H}{\partial t}+u\frac{\partial H}{\partial r}+H\frac{\partial u}{\partial r}=0 \, , \tag{4}$$

where $\rho\,(r,t)$, $u\,(r,t)$, $q\,(r,t)$, $H\,(r,t)$, $p\,(r,t)$, $t$ and $r$ are density, particle velocity, viscous stress tensor, axial magnetic field, pressure, time coordinate and position coordinate with respect to the origin in the direction normal to the shock front, respectively, and $\mu'$ being the constant magnetic permeability of the gas taken to be unity throughout the problem. It is to be noted that the diffusion term is omitted in the Equation (3) by virtue of the assumed perfect electrical conductively. The viscous stress tensor $q$ is given by

$$q=(4/3)\mu(du/dr) \tag{5}$$

where, $\mu$ is the coefficient of viscosity. For simplicity, it is assumed that $\mu$ is independent of temperature. It is remarkable that with conditions $H_\theta=H_r=0$ and $H_z=H\neq 0$, Equation 4 can be written as $\partial\boldsymbol{H}/\partial t+\nabla\times(\boldsymbol{H}\times\boldsymbol{u})=0$ and further $\partial(\nabla\cdot\boldsymbol{H})/\partial t=0$. Thus, the Maxwell equation $\nabla\cdot\boldsymbol{H}=0$ is included in Equation 4.

In a coordinate system with stationary shock front, the shock strength remains practically unchanged during the small time interval $\Delta t$ required to travel a distance of the order of the shock front thickness, as a result the term containing the partial derivative with respect to time ( $\partial/\partial t$ ) is dropped in the Equations (1) to (4) and further the partial derivative ( $\partial/\partial r$ ) is replaced by the total derivative ( $d/dr$ ). Thus, the Equations (1) to (4) governing the one dimensional plane symmetrical flow of a viscous gas under the influence of an axial magnetic field are written as

$$u\frac{d\rho}{dr}+\rho\frac{du}{dr}=0 \, , \tag{6}$$

$$\frac{d(\rho u^2 + P - q)}{dr} + \frac{1}{2}\frac{dH^2}{dr} = 0\,, \tag{7}$$

$$\frac{d[\rho u(E + u^2/2) + pu - qu]}{dr} = 0\,, \tag{8}$$

$$u\frac{dH}{dr} + H\frac{du}{dr} = 0\,. \tag{9}$$

where $\gamma$ is the adiabatic index, i.e., $\gamma = C_p/C_v$. The boundary condition on the solution of these differential Equations (6) to (9) requires that the gradient of flow variables must vanish ahead of the shock front (at $r = +\infty$) as well as behind the shock front (at $r = -\infty$). With these limits, the initial flow variables designated by the subscript '$o$' are $p_o, \rho_o, u_o, H_o$ and the final flow variables with no subscript are $p$, $\rho$, $u$, $H$. If the shock front is moving with velocity $U$, then in the coordinate system fixed with the shock front, the initial particle velocity $u_o$ will be

$$u_o = U \tag{10}$$

**3. Exact Solutions for the flow variables**

In order to obtain the exact solutions for the flow variables, we need to solve the flow Equations (6) to (9) using the boundary condition given by Equation (10) in the equilibrium state. For this, we integrate the Equations (6) to (9) which yields,

$$\rho = \rho_0 U/u \tag{11}$$

$$p = p_o + q + \rho_o U^2 - \rho u^2 - H^2/2 + H_o{}^2/2 \tag{12}$$

$$p\,u\gamma/(\gamma-1) + \rho u^3/2 - qu = p_o U\gamma/(\gamma-1) + \rho_o U^3/2 \tag{13}$$

$$H = H_o U/u \tag{14}$$

Now using Equations (11) to (12) and (14), the Equation (13) becomes

$$\begin{aligned}\gamma\, p_o\, u + \gamma\, qu + \gamma\rho_o U^2 u + \gamma\, H_o^2 u/2 - \gamma\rho_o U u^2 - \gamma\, H_o^2 U^2/2u + (\gamma-1)\rho_o U u^2/2 - (\gamma-1)qu \\ = \gamma\, p_o U + (\gamma-1)\rho_o U^3/2\end{aligned} \tag{15}$$

Let us introduce two new dimensionless quantities called particle velocity $\eta$ and shock strength $M$ as

$$\eta = u/U \text{ and } M = U/a_o \tag{16}$$

where $a_o$ is the speed of sound in the unperturbed state. Using Equations (5) and (16) in Equation (15), we get

$$\begin{aligned}[(\gamma+1)/2]\eta^3 - (\gamma + M^{-2} + H_o^2/2p_o M^2)\eta^2 + (M^{-2} + (\gamma-1)/2)\eta + H_o^2/2p_o M^2 \\ = [(4/3)\mu/M(\gamma\, p_o\, \rho_o)^{1/2}]\eta^2 d\eta/dr\end{aligned} \tag{17}$$

Thus, the Equation (17) can be written as

$$a\eta^3 + 3b\eta^2 + 3c\eta + d = e\eta^2 d\eta / dr \quad (18)$$

Where

$$a = (\gamma+1)/2, \quad b = -(\gamma + M^{-2} + H_o^2 / 2p_o M^2)/3, \quad c = (M^{-2} + (\gamma-1)/2)/3,$$

$$d = H_o^2 / 2p_o M^2 \text{ and } e = [(4/3)\mu / M(\gamma\ p_o\ \rho_o)^{1/2}]$$

Since outside the transition region, there is no gradient in the flow variables in the equilibrium state. Therefore, in the equilibrium state, we can write

$$d\eta / dr = 0 \text{ with } \eta = \eta_{eq}$$

With this equilibrium condition, the Equation (18) becomes a cubic equation with respect to the particle velocity in equilibrium state $\eta_{eq}$. For searching real solutions, we put cubic equation in the form as

$$Z^3 + 3FZ + G = 0 \quad (19)$$

where

$$Z = a\,\eta_{eq} + b,\ F = a\,c - b^2 \text{ and } G = a^2\ d - 3a\,b\,c + 2\,b^3$$

Now defining, $\tan\phi = -K/G$, where $G^2 + 4F^3 = -K^2$. The Equation 19 is solved using the Cordom's method and thus, the algebraic solutions are

$$\eta_{eq} = \eta_1 = 2(-F)^{1/2}\cos(\phi/3),\ \ \eta_{2,3} = -2(-F)^{1/2}\cos[(\pi \pm \phi)/3] \quad (20)$$

The three roots, given by Equation (20) will be real if the condition $G'^2 + 4F'^3 < 0$ is satisfied. Using this condition and Equation (20), we can numerically compute the particle velocity corresponding to the equilibrium state in which there are no gradients in the flow variables. Let us choose the origin at the point of inflection of the velocity profile. The point of inflection is obtained by using, the condition $d^2\eta / dr^2 = 0$ into the Equation (18) which again yields a cubic equation given as

$$\eta_{in}^3 + 3F'\eta_{in} + G' = 0 \quad (21)$$

where

$$F' = -c/3a,\ G' = -2d/a$$

Defining, $\tan\phi' = -K'/G'$, where $G'^2 + 4F'^3 = -K'^2$. Using the Cordon's method, the algebraic solutions of the Equation (21) are

$$\eta_{in} = \eta_1' = 2(-F')^{1/2}\cos(\phi'/3), \eta_{2,3}' = -2(-F')^{1/2}\cos[(\pi \pm \phi')/3] \quad (22)$$

The three roots, given by the Equation (22), will be real if the condition $G'^2 + 4F'^3 < 0$ is satisfied. Using this condition and Equation (22), we can determine the point of inflection at the velocity profile. On integration the Equation (18) yields an analytic solution for *r* as

$$r = (e/a)[A\log\{(\eta-\eta_1)/(\eta'_{in}-\eta_1)\} + B\log\{(\eta-\eta_2)/(\eta'_{in}-\eta_2)\} + C\log\{(\eta-\eta_3)/(\eta'_{in}-\eta_3)\}] \qquad (23)$$

where

$$A = \eta_1^2/(\eta_1-\eta_2)(\eta_1-\eta_3),\ B = \eta_2^2/(\eta_2-\eta_3)(\eta_2-\eta_1) \text{ and } C = \eta_3^2/(\eta_3-\eta_1)(\eta_3-\eta_2)$$

Equation (23) gives a relation between the particle velocity and the distance *r*. Thus, the particle velocity depends on the distance *r* within the shock transition region. Using Equations (13) and (18), we can write the temperature across the shock front as

$$T/T_o = 1 + (\gamma\text{-}1)M^2[\gamma\,\eta^2/2 - (\gamma + M^{-2} + H_o^2/2p_oM^2)\eta + (M^{-2} + \gamma/2) + H_o^2/2p_oM^2\eta] \qquad (24)$$

The Equation (24) gives a relation between the temperature and the particle velocity and it is obvious from Equation (23) that the particle velocity depends on the distance *r*. Hence using the Equation (24) we can study the variations of the temperature with respect to the distance *r* within the shock transition region. Using the Equations (5), (11), (14) and (18), we can write the expression for the pressure across the shock front as

$$p/p_o = 1 + (1-\eta)\gamma\,M^2 + (1-\eta^{-2})H_o^2/2p_o + \gamma\,M^2(a\eta^3 + 3b\eta^2 + 3c\eta + d)/\eta^2 \qquad (25)$$

This Equation (25) gives a relation between the pressure and the particle velocity, and from Equation 23 it is obvious that the particle velocity depends on the distance *r*. Thus, using Equation (25) we can study the variations of the pressure with distance *r* within the shock transition region.

Further, the change-in-entropy across the shock front is given by

$$(\Delta S/R)_\eta = \gamma(\gamma-1)^{-1}\log(T/T_o) - \log(p/p_o) \qquad (26)$$

Thus, the entropy production across the MHD shock front is easily obtained by substituting Equations (24) and (25) into Equation (26).

**4. Results and Discussions**

In the present paper, an exact solution was obtained for MHD shock waves in a viscous gas. The general non-dimensional forms of the analytical expressions for the distribution of the flow variables (the particle velocity $\eta$, the temperature $T/T_o$, the pressure $p/p_o$, and the change-in-entropy $\Delta S/R$) within the shock transition region are given by Equations (23) to (26), respectively. These analytical expressions were derived by assuming that the disturbances due to the reflections, wave interactions in the wake, etc., do not overtake the shock waves. The general non-dimensional analytical expressions for the particle velocity, temperature, pressure and change-in-entropy are the functions of the distance $r$, coefficient of viscosity $\mu$, strength of magnetic field $H_o$, strength of shock $M$ and adiabatic index $\gamma$ of the gas. Therefore, the values of the constant parameters are taken to be $\mu = 15\times10^{-6}$, $17.2\times10^{-6}$, $20\times10^{-6}$ pascal.sec, $H_O = 0,\ 0.2,\ 0.4,\ 0.6,\ 0.8$, $M = 1.1,\ 1.5,\ 2.0$, $\gamma = 1.33,\ 1.40,\ 1.66$, initial pressure $p_O = 0.9,\ 1.0,\ 1.1$ bar and initial density $\rho_O = 1.20, 1.29, 1.40$ Kg/m$^3$ for the general purpose of numerical computations.

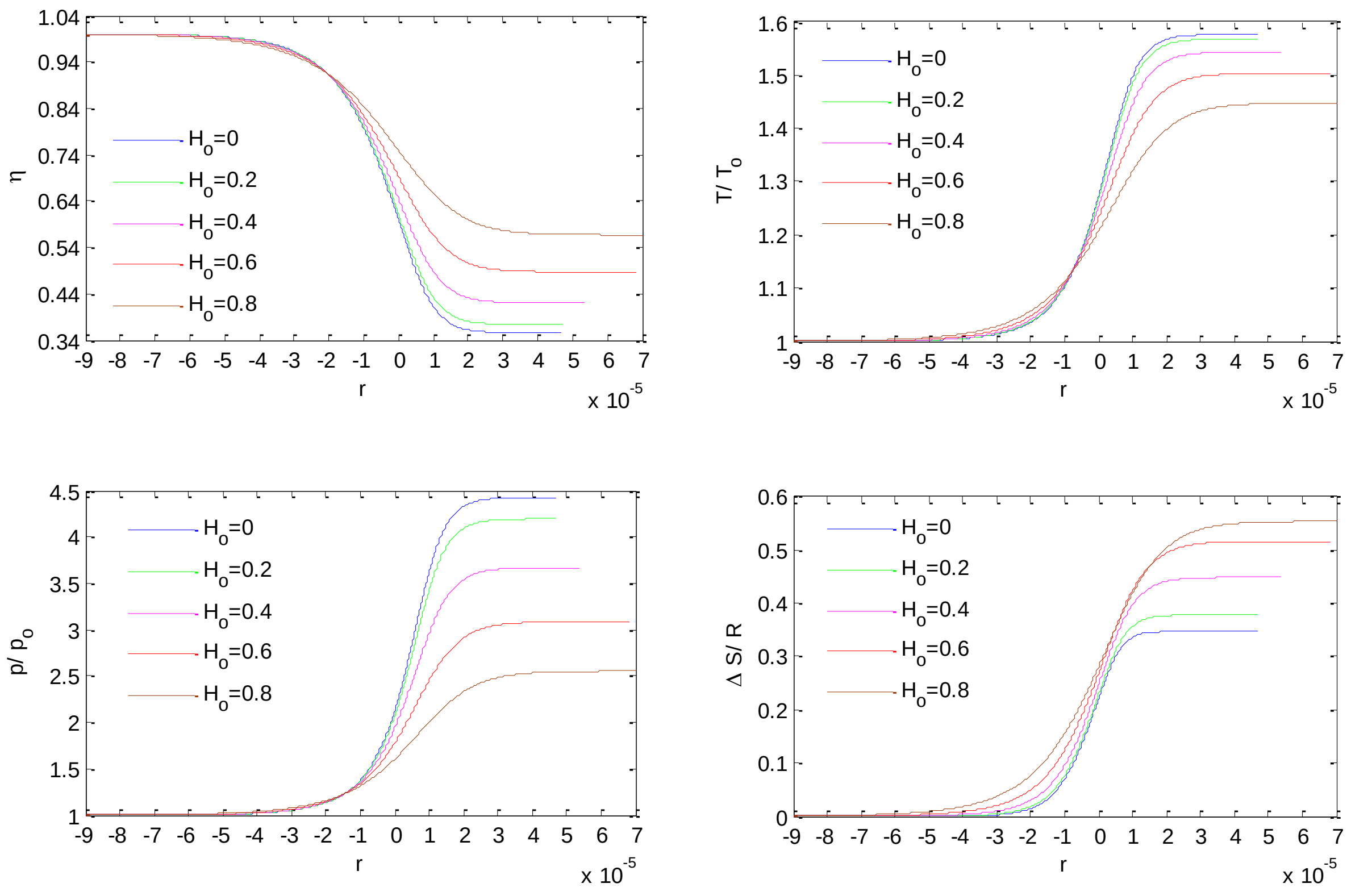

**Fig. 1** Non-dimensional particle velocity, temperature, pressure and change-in-entropy distribution with distance for various values of $H_o$ and constant values of $M = 2$, $\gamma = 1.33$, $p_o = 0.9$bar, $\rho_o = 1.2$Kg/m$^3$ and $\mu = 15\times10^{-6}$ pascal.sec.

Fig. 1 shows the variations of the particle velocity, the temperature, the pressure and the change-in-entropy with respect to the propagation distance for the constant values of $M = 2$, $\gamma = 1.33$, $p_o = 0.9$bar $\rho_o = 1.20$ Kg/m$^3$ $\mu = 15\times10^{-6}$ pascal.sec and different values of axial magnetic field $H_o = 0, 0.2, 0.4, 0.6, 0.8$. It is notable that for small values of magnetic field the spreading of the flow variables is lesser than that for large values of magnetic field. However, the effect of increase in the strength of magnetic field on the spreading of flow variables is appreciable ahead of the point of inflection. Therefore, the presence of magnetic field increases the thickness of shock front and it is observed that the thickness is maximum corresponding to the highest value of the strength of magnetic field. Thus, the thickness of MHD shock front increases with increasing strength of the magnetic field. It is also observed that the range of variations of the flow variables decreases with increasing magnetic field.

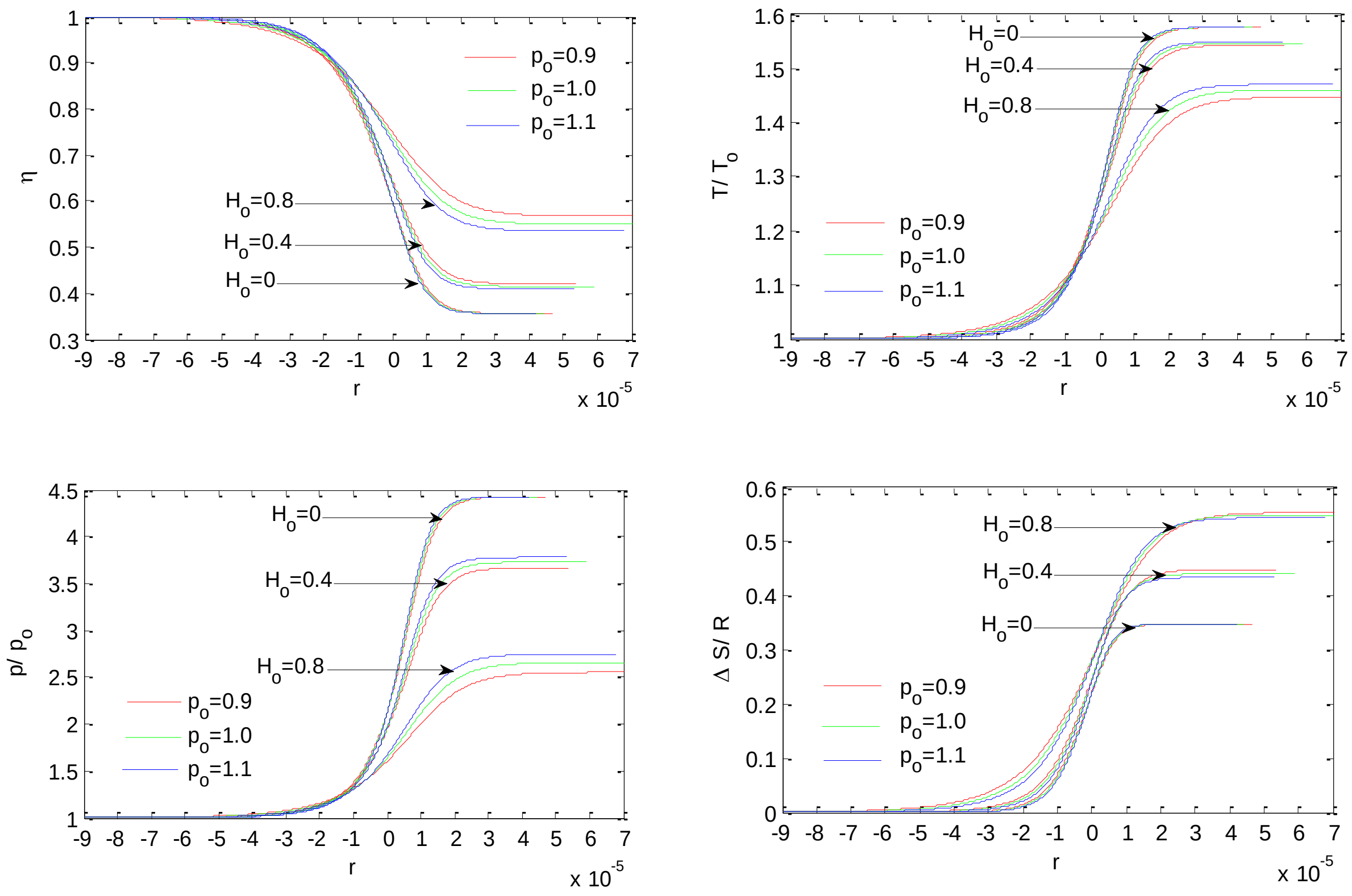

**Fig. 2** Non-dimensional particle velocity, temperature, pressure and change-in-entropy distribution with distance for various values of $p_o$ and $H_o$ and constant values of $M = 2$, $\gamma = 1.33$, $\rho_o$ = 1.2 Kg/m$^3$ and $\mu = 15\times10^{-6}$ pascal.sec.

Fig. 2 shows the variations of the particle velocity $\eta$, the temperature $T/T_o$, the pressure $p/p_o$ and the change-in-entropy $\Delta S/R$ with distance for the different values of initial pressure $p_o = 0.9, 1, 1.1$ bar and axial magnetic field $H_o = 0, 0.4, 0.8$ Tesla and constant values of $M = 2$, $\gamma = 1.33$, $\rho_o = 1.2$ kg/m$^3$, $\mu = 15\times10^{-6}$ pascal.sec. The spreading of the flow variables decreases with increase in the value of initial pressure. However, the decrease in the thickness of shock front with increase in the initial pressure is more noticeable for large values of the strength of magnetic field than that for small values of the strength of magnetic field. Thus, the thickness of MHD shock front increases with decrease in the initial pressure.

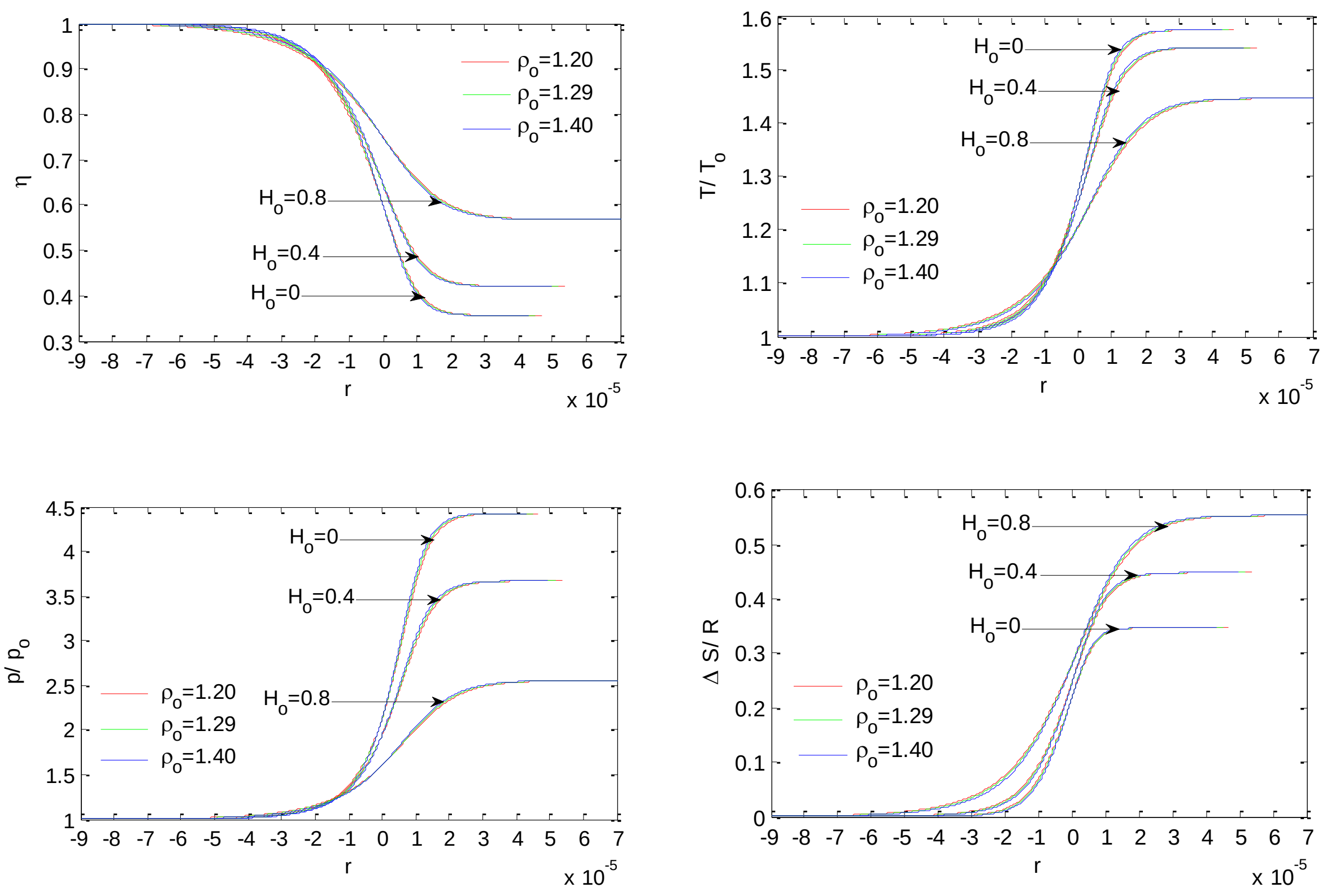

**Fig. 3** Non-dimensional particle velocity, temperature, pressure and change-in-entropy distribution with distance for various values of $\rho_o$ and $H_O$ and constant values of $M = 2$ , $\gamma = 1.33$ , $p_o$ = 0.9bar and $\mu = 15\times10^{-6}$ pascal.sec.

Fig. 3 shows the variations of the particle velocity $\eta$ , the temperature $T/T_o$ , the pressure $p/p_o$ and the change-in-entropy $\Delta S/R$ with distance for the different values of the magnetic field $H_o = 0, 0.4, 0.8$ Tesla and the initial density $\rho_o = 1.2, 1.29, 1.4$ Kg/m$^3$ and constant values of $M = 2$ , $\gamma = 1.33$ , $p_o$ =0.9bar and $\mu = 15\times10^{-6}$ pascal.sec. It is notable that the spreading of the flow variables decreases with increase in the value of initial density. Thus, the thickness of MHD shock front increases with decreasing initial density. However, the decrease in the thickness of shock front with increase in the initial density is independent of the strength of magnetic field.

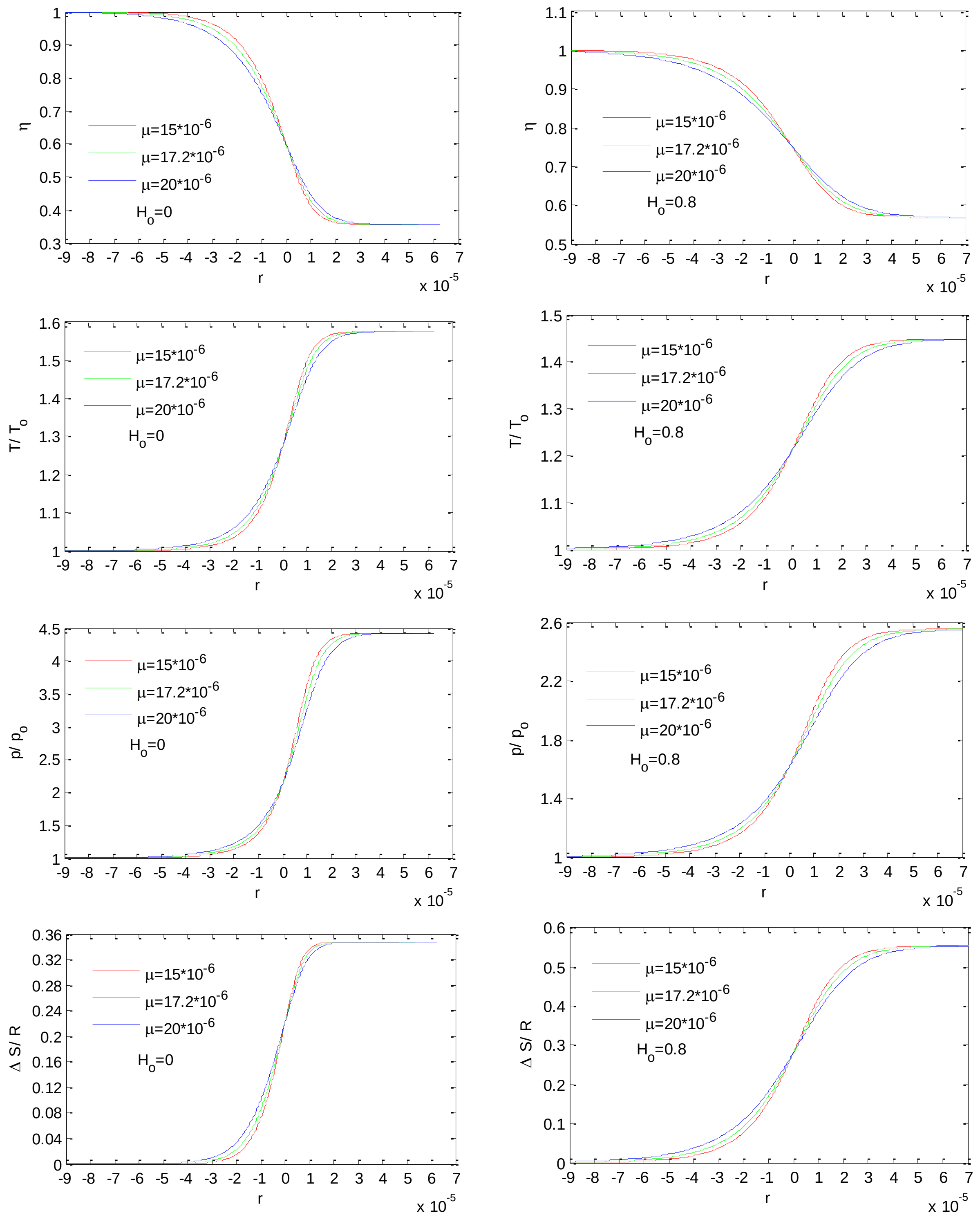

**Fig. 4** Non-dimensional particle velocity, temperature, pressure, and change-in-entropy distribution with distance for various values of $\mu$ and $H_o$ and constant values of $M = 2$, $\gamma = 1.33$, $\rho_o = 1.20$ Kg/m$^3$ and $p_o$ = 0.9 bar.

Fig. 4 shows the variations of the particle velocity $\eta$, the temperature $T/T_o$, the pressure $p/p_o$ and the change-in-entropy $\Delta S/R$ with distance $r$ for different values of the magnetic field $H_o = 0, 0.8$ Tesla and the coefficient of viscosity $\mu_o = 15\times10^{-6}, 17.2\times10^{-6}, 20\times10^{-6}$ pascal.sec. It is noteworthy that the spreading of the flow variables increases with increase in the value of the coefficient of viscosity. Thus, the thickness of MHD shock front increases with increase in the value of the coefficient of viscosity. However, the increase in the thickness of shock front with increase in the viscosity coefficient is more noticeable for large values of the strength of magnetic field.

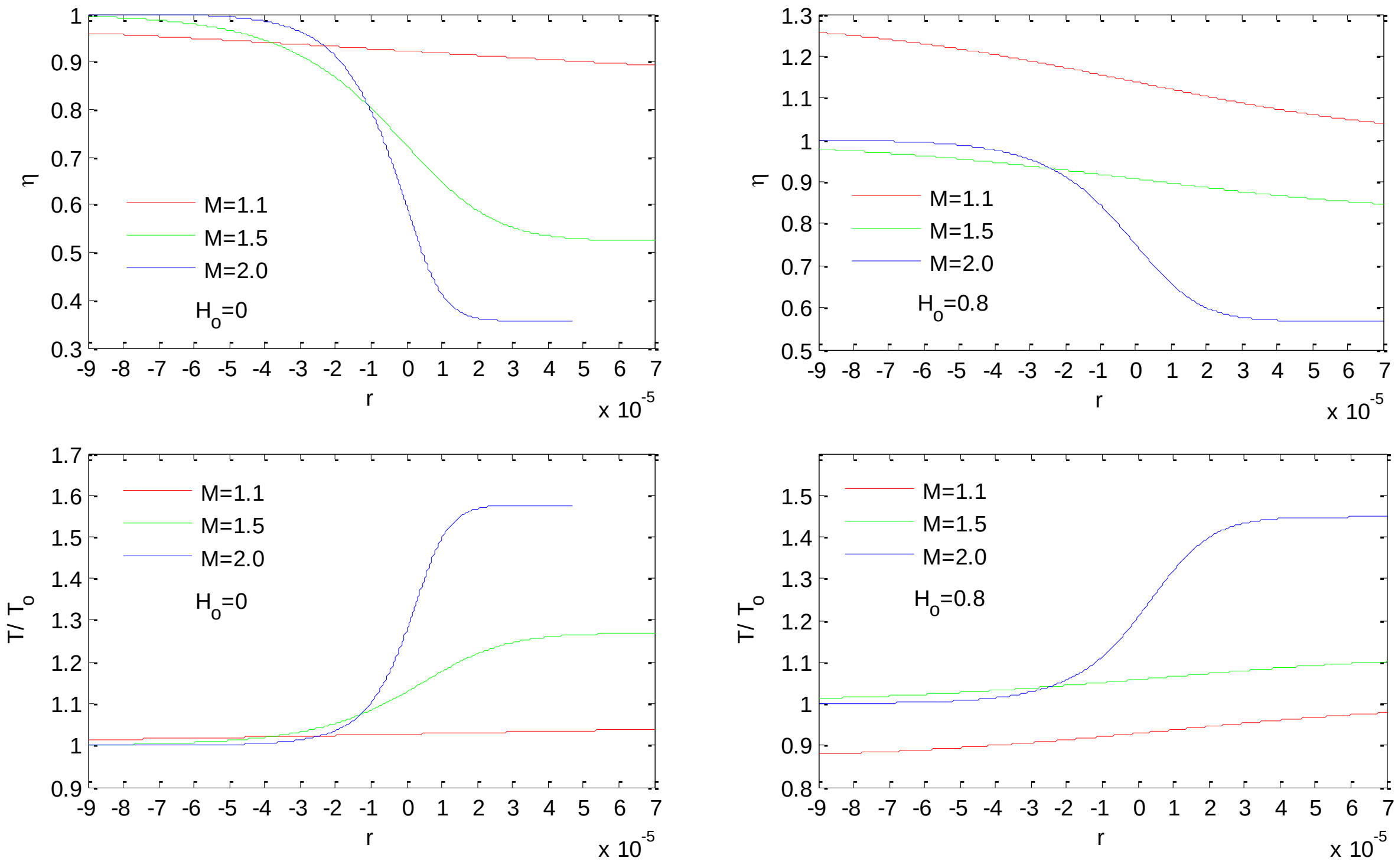

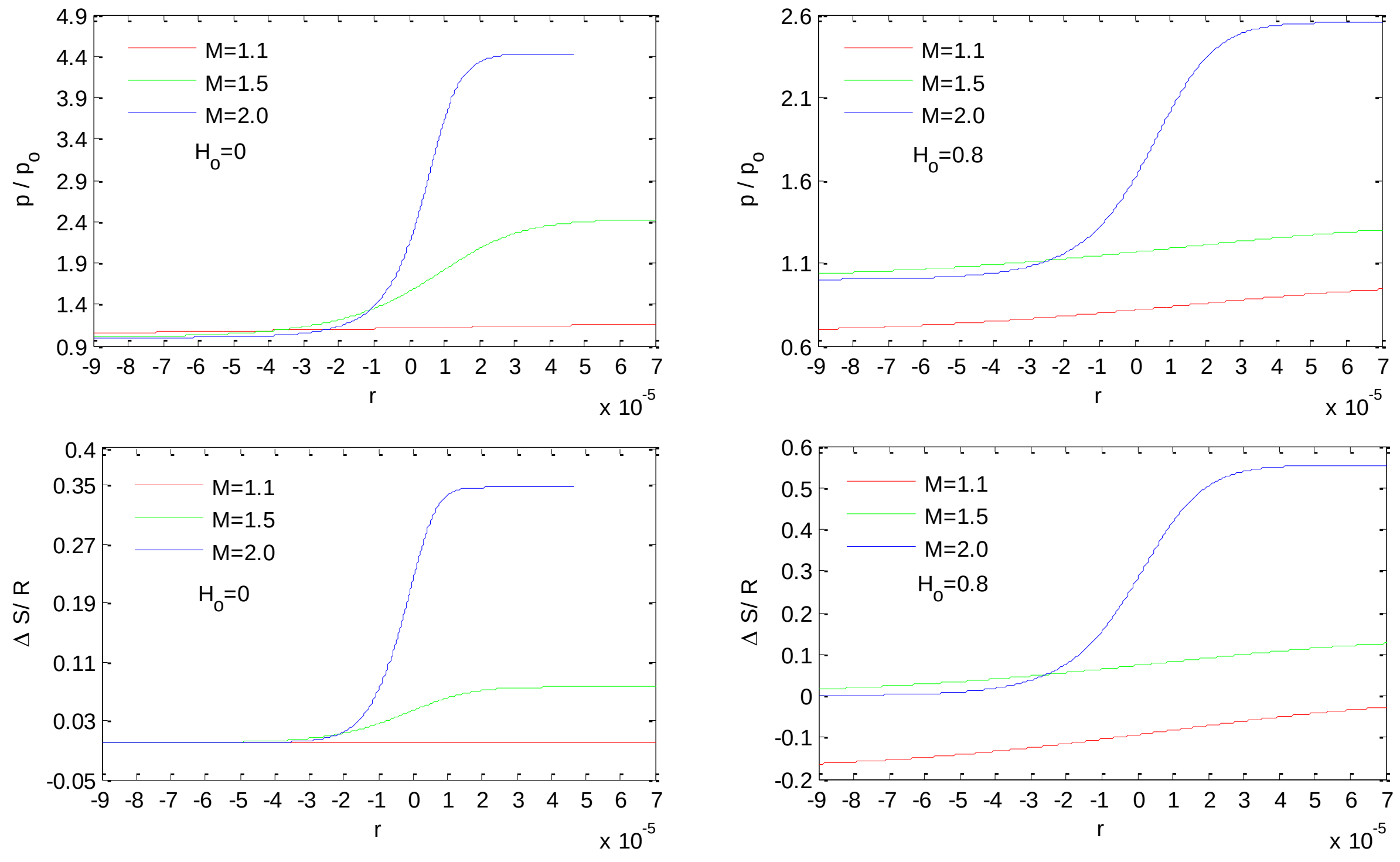

**Fig. 5** Non-dimensional particle velocity, temperature, pressure and the change-in-entropy distribution with distance for various values of $M$ and $H_o$ and constant values of $\gamma = 1.33$, $\rho_o = 1.2$ Kg/m$^3$, $\mu = 15\times10^{-6}$ pascal.sec and $p_o = 0.9$ bar.

Fig. 5 shows the variations of the particle velocity $\eta$, the temperature $T/T_o$, the pressure $p/p_o$ and the change-in-entropy $\Delta S/R$ with distance for different values of the magnetic field $H_o = 0, 0.8$ Tesla and the shock strength $M = 2, 5, 10$ and constant values of $\gamma = 1.33$, $\rho_o = 1.2$ Kg/m$^3$, $\mu = 15\times10^{-6}$ pascal.sec and $p_o = 0.9$ bar. It is remarkable that for small values of shock strength the spreading of the flow variables is more than that for large values of the shock strength. Thus, the thickness of MHD shock front decreases with increasing strength of the shock wave. However, the change in the spreading of the flow variable due to the change in shock strength is more for large values of the strengths of magnetic field. It is to be noted that the effect of magnetic field is appreciable for small values of shock strength, while for large values of shock strength it is very small.

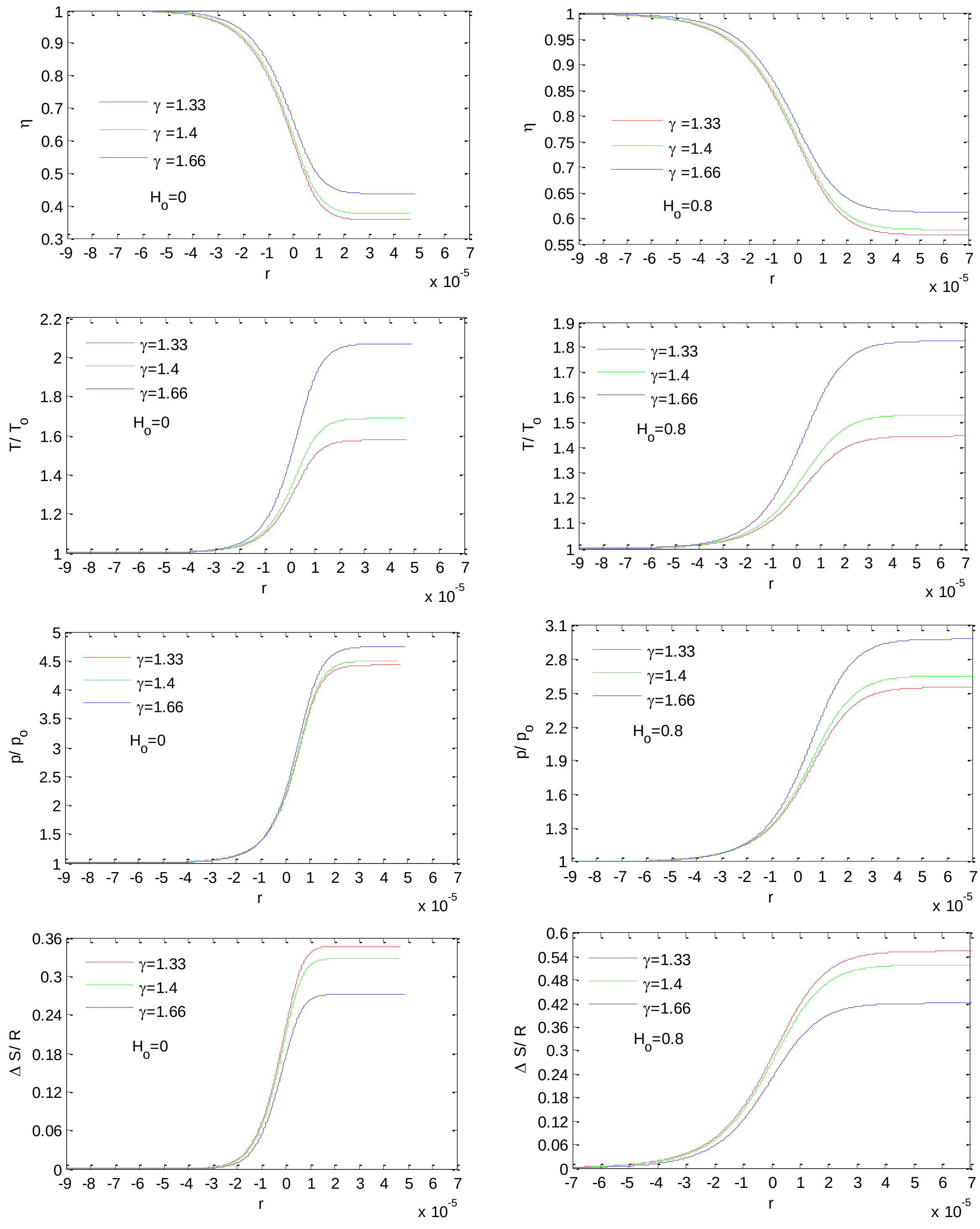

**Fig. 6** Non-dimensional particle velocity, temperature, pressure and change-in-entropy distribution with distance for various values of $\gamma$ and $H_o$ and constant values of $M = 2$, $\rho_o = 1.2$ Kg/m$^3$, $\mu = 15\times10^{-6}$ pascal.sec and $p_o = 0.9$ bar.

Fig. 6 shows the variations of the particle velocity $\eta$, the temperature $T/T_o$, the pressure $p/p_o$ and the change-in-entropy $\Delta S/R$ with distance for different values of the adiabatic index $\gamma = 1.33, 1.4, 1.66$ and magnetic field $H_o = 0,\ 0.8$ Tesla and for constant values of $M = 2$, $\rho_o$ = 1.2 Kg/m$^3$, $\mu = 15\times10^{-6}$ pascal.sec and $p_o = 0.9$ bar. It is notable that the spreading of the flow variables increases ahead of inflection point and decreases behind the inflection point with increase in the value of adiabatic index. Thus, the thickness of MHD shock front increases with increase in the value of adiabatic index. However, the change in the spreading of the flow variables with the adiabatic index is independent of the strength of magnetic field.

**5. Conclusions**

The investigations made in the present paper are intended to contribute to the understanding of the structure of MHD shock waves in a viscous gas, by giving, for the first time, the full exact solutions for the flow field within the shock transition region. The analysis presented in the paper shows the fundamental role played by viscosity in determining MHD shock structure. The following conclusions may be drawn from the findings of the current analysis:

1. The thickness of MHD shock front increases with increasing strength of the magnetic field. The influence of magnetic field on the spreading of the flow variables is appreciable ahead of the point of inflection.

2. The thickness of MHD shock front increases with decreasing initial pressure and the change in thickness is more noticeable for large values of the strength of magnetic field than that for small values of the strength of magnetic field.

3. The thickness of MHD shock front increases with decreasing initial density and the change in thickness with initial density is independent of the strength of magnetic field.

4. The thickness of MHD shock front increases with increase in the viscosity and the change in thickness is more noticeable for large values of the strength of magnetic field.

5. The thickness of MHD shock front decreases with increasing strength of the shock wave and the change in thickness is more for large values of the strengths of magnetic field. It is to be noted that the effect of magnetic field is appreciable for small values of shock strength, while for large values of shock strength it is very small.

6. The thickness of MHD shock front increases with increase in the value of adiabatic index and the change in thickness with the adiabatic index is independent of the strength of magnetic field.

Thus, the results obtained here are important for the cases where the viscosity of fluid plays an important role in addition to the various industrial and mechanical engineering processes where ferro fluids are used. The findings may also be useful in the study of the effects of magnetic fields on blood circulation, cardiovascular events, crude oil transportation, etc.